\documentclass[aps,onecolumn,showpacs,nofootinbib,showkeys]{revtex4-2}
\usepackage[margin=2.15cm]{geometry}

\RequirePackage[T1]{fontenc}

\usepackage{epsfig,graphicx,amsmath,amssymb,bm}
\RequirePackage{mathptmx}      
\usepackage{mathrsfs}
\usepackage{amssymb}
\RequirePackage{color}
\usepackage{changes}
\usepackage{slashed}
\usepackage{multirow}
\usepackage{scalerel}
\usepackage{tikz-feynman}
\usepackage{textcomp}
\usepackage{subcaption}
\usepackage[english]{babel}
\usepackage{float}
\RequirePackage{hyperref}
\hypersetup{
    linktocpage,
    colorlinks,
    citecolor=blue,
    filecolor=black,
    linkcolor=blue,
    urlcolor=blue,
}
\usepackage{nccmath}

\begin{document}

\title{Decays $\tau \to f_0(\pi,K) \nu_\tau$ and $\tau \to 3 \pi \nu_\tau$ accounting for the contribution of $f_0(500)$}


\author{M.K. Volkov$^{1}$}\email{volkov@theor.jinr.ru}
\author{A.A. Pivovarov$^{1}$}\email{pivovarov@theor.jinr.ru}
\author{K. Nurlan$^{1,2}$}\email{nurlan@theor.jinr.ru}

\affiliation{$^1$ Bogoliubov Laboratory of Theoretical Physics, JINR, 
                 141980 Dubna, Moscow region, Russia \\
                $^2$ The Institute of Nuclear Physics, Almaty, 050032, Kazakhstan}   


\begin{abstract} 
In the $U(3) \times U(3)$ quark NJL model, $\tau$ lepton decays with the production of scalar mesons $f_0(\pi,K])$ and neutrinos are studied, where $f_0=f_0(500), f_0(980)$. It is shown that these decays mainly occur via contact channels and channels with axial-vector mesons $a_1$, $K_1(1270)$ and $K_1(1400)$. All mesons are considered as quark-antiquark states in this case. The obtained results can be considered as predictions for future experiments. The obtained estimates for the branching fractions of the $\tau \to \pi^-2\pi^0 \nu_\tau$ decay taking into account the contributions of the $\rho\pi$ and $f_0(500)\pi$ states are in satisfactory agreement with experimental data.

\end{abstract}


\maketitle

\section{Introduction}
The Nambu-Jona-Lasinio model is a good tool for describing the internal properties and interactions of mesons at low energies \cite{Nambu:1961tp,Eguchi:1976iz,Ebert:1982pk,Volkov:1984kq,Volkov:1986zb,Ebert:1985kz,Vogl:1991qt,Klevansky:1992qe,Volkov:1993jw,Ebert:1994mf,Hatsuda:1994pi,Buballa:2003qv,Volkov:2005kw}. In this model, all mesons are considered as quark-antiquark systems. However, when describing the mass of the isovector scalar meson $a_0(980)$, difficulties arise due to the fact that tetraquark states, kaonic molecules or more complex structures should be taken into account \cite{Volkov:1998ax,Volkov:1999yi,Ebert:2000nx,Volkov:2001ct,Achasov:1999wv,tHooft:2008rus,Kim:2017yvd}. 
The masses and interactions of the isoscalar mesons $f_0(500)$, $f_0(980)$ and the strange meson $K^*_0(800)$ are described in satisfactory agreement with experiment based on the assumption that quark-antiquark structures are the main states of these mesons \cite{Volkov:1998ax,Volkov:1999yi,Ebert:2000nx,Volkov:2001ct}.

In recent works \cite{Volkov:2022ukj,Volkov:2022nxp}, the decays of the $\tau$ lepton with strange scalar mesons $\tau \to K^*_0(800)[K, \pi, \eta] \nu_\tau$, $\tau \to K^*_0(1430) \pi \nu_\tau$ and the main decays of the meson $K^*_0(800)$ in the ground and excited states were described in the NJL model where the decay widths were obtained in agreement with experiment.

The present work is devoted to the description of tau lepton decays with the production of scalar mesons $\tau \to f_0(500)[\pi,K] \nu_\tau$ and $\tau \to f_0(980)[\pi,K] \nu_\tau$. Note that calculations performed in this paper lead to a large value of the width $\tau \to f_0(500)\pi \nu_\tau$. However, despite this, direct experimental data related to this decay are absent. The absence of experimental data for $\tau \to f_0(500)\pi \nu_\tau$ can be explained by the large width of the decay $f_0(500) \to 2\pi$ ($\Gamma_{f_0(500)} = 100-800 $ MeV) \cite{ParticleDataGroup:2024cfk}. The role of the intermediate channel with the scalar meson $f_0(500)$ was experimentally investigated in the decay $\tau \to \pi^-2\pi^0 \nu_\tau$ \cite{CLEO:1999rzk}. The experiment showed that the individual contribution of the $f_0(500)$ channel to the total decay width is about 15\%.

It is interesting to note that the $\tau \to f_0(980)\pi \nu_\tau$ decay was investigated in \cite{Dai:2018rra} using the triangular singularity mechanism.

\section{Effective quark-meson Lagrangian of the NJL model}
In this paper, we will use the effective Lagrangian $U(3) \times U(3)$ of the NJL quark model that describes the interactions of quarks with mesons obtained after bosonization of the local four-quark chiral Lagrangian. The derivation of the quark-meson Lagrangian based on the initial four-quark Lagrangian of the NJL model (\ref{Lagrangian}) is described in the works \cite{Volkov:1986zb,Volkov:1993jw,Volkov:2005kw}. Note that it is necessary to take into account the singlet-octet mixing when describing the mesons $f_0(500)$ and $f_0(980)$. In the NJL model, this can be described by introducing the 't Hooft interaction, which leads to mixing of states containing light $u,d$-quarks with a heavier $s$-quark one \cite{Volkov:1998ax,Volkov:1999yi,Ebert:2000nx,Volkov:2001ct}. Accounting for the 't Hooft interaction allows one to correctly describe the masses of scalar mesons $f_0(500)$ and $f_0(980)$. A similar situation occurs when describing the masses of the pseudoscalar mesons $\eta$ and $\eta'$. As a result, the quark-meson Lagrangian of the interaction of mesons with quarks has the form
	\begin{eqnarray}
	\label{Lagrangian}
		\Delta \mathcal{L}_{int} & = &
		\bar{q} \left[ \sum_{j = \pm, 0}\left[i g_{\pi} \gamma^{5} \lambda_{j}^\pi \pi^{j}  +
        i g_{K} \gamma^{5} \lambda_{j}^K K^{j} + \frac{g_{a_1}}{2}\gamma^{\mu}\gamma^{5}\lambda^{a_1}_{j}a^{j}_{1\mu}\right] + 
        \right. \nonumber\\
	&&\left.
        \sum_{j = \pm, 0}
        \left[ \frac{g_{K_1}}{2}\gamma^{\mu}\gamma^{5}\lambda^{K}_{j}K^{j}_{1A\mu} +
        \frac{g_{\rho}}{2}\gamma^{\mu}\lambda^{\rho}_{j}\rho^{j}_{\mu} +
         g_{\sigma_u}\lambda^{a_0}_{j} {a_0}^{j}
        \right] +
        \right. \nonumber\\
	&&\left.
        g_{{f_0}_u} \lambda_u {f_0}_u + g_{{f_0}_s} \lambda_s {f_0}_s +
        ig_{\eta_u} \lambda_u \eta_u + ig_{\eta_s} \lambda_s \eta_s
        \right]q,
	\end{eqnarray}
   where $q$ and $\bar{q}$ are the $u$, $d$, and $s$ quarks triplet; $\lambda$ are linear combinations of the Gell-Mann matrices. Isoscalar mesons are mixed states
        \begin{eqnarray}
        {f_0}_u = f_0(500) \cos{\bar{\phi}} + f_0(980) \sin{\bar{\phi}}, \\
        {f_0}_s = -f_0(500) \sin{\bar{\phi}} + f_0(980) \cos{\bar{\phi}}, \nonumber
	\end{eqnarray}
    
	\begin{eqnarray}
        \eta_u = -\eta \sin{\bar{\theta}} + \eta' \cos{\bar{\theta}}, \\
        \eta_s = \eta \cos{\bar{\theta}} + \eta' \sin{\bar{\theta}}, \nonumber
	\end{eqnarray}
    where the meson mixing angles $\bar{\phi} = \phi - \phi_0$, $\bar{\theta} = \theta - \theta_0$. $\theta= -19^\circ$ and $\phi= 24^\circ$ are deviations from the ideal mixing angle $\theta_0=\phi_0\approx 35.3^\circ$ \cite{Volkov:1998ax,Ebert:2000nx}.

        The coupling constants of mesons with quarks are
	\begin{eqnarray}
	\label{Couplings}
        g_{{f_0}_u} = g_{\sigma_u} = \left(4I_{20}\right)^{-1/2}, 
        \quad g_{{f_0}_s} = \left(4I_{02}\right)^{-1/2},
        \quad g_{\eta_u} = g_\pi = \left(\frac{4}{Z_{\pi}}I_{20}\right)^{-1/2},
        \quad g_{\eta_s} = \left(\frac{4}{Z_{s}}I_{02}\right)^{-1/2},
	\end{eqnarray}
    
    \begin{eqnarray}
	g_{a_1} = g_{\rho} = \left(\frac{2}{3}I_{2}\right)^{-1/2},
        \quad g_{K} =  \left(\frac{4}{Z_{K}}I_{11}\right)^{-1/2},
        \quad g_{K_1} = \left(\frac{2}{3}I_{11}\right)^{-1/2},
	\end{eqnarray}    
    where $Z_{\pi}$, $Z_s$ and $Z_K$ are additional renormalization constants appearing in transitions between pseudoscalar and axial-vector mesons
	\begin{eqnarray}
	Z_{\pi} = \left[1 - 6\frac{m^2_u}{M_{a_{1}}^{2}}\right]^{-1},
        \quad Z_K = \left[1 - \frac{3(m_s+m_u)^2}{2M_{K_{1A}}^{2}} \right]^{-1},
        \quad Z_s = \left[1 - 6\frac{m^2_s}{M_{f_1(1420)}^{2}}\right]^{-1},
	\end{eqnarray}
    where the constituent quark masses are $m_{u} \approx m_{d} = m = 270$~MeV, $m_s = 420$~MeV arising as a result of spontaneous breaking of chiral symmetry ~\cite{Volkov:2005kw}; $M_{K_{1A}}$ is the effective mass appearing as a result of mixing strange axial-vector meson states $K_1(1270)$ and $K_1(1400)$ 
        \begin{eqnarray}
	M_{K_{1A}} = \left[ \frac{\sin^2(\alpha)}{M^2_{K_1(1270)}} + \frac{\cos^2(\alpha)}{M^2_{K_1(1400)}} \right]^{-1},
	\end{eqnarray}
    where $\alpha=57^{\circ}$ \cite{Volkov:1984gqw,Suzuki:1993yc,Volkov:2019awd}.
    
Integrals appearing in the quark loops are
 	\begin{eqnarray}
	\label{int}
		I_{nm} =
		-i\frac{N_{c}}{(2\pi)^{4}}\int\frac{\Theta(\Lambda_q^{2} + k^2)}{(m^2 - k^2)^n(m^2_s - k^2)^m} \mathrm{d}^{4}k, 
	\end{eqnarray}
    where $\Lambda_q = 1265$~MeV is the cutoff parameter and $N_{c} = 3$ is the number of colors in QCD.

\section{Decays $\tau \to \pi^- f_0 \nu_{\tau}$ and $\tau \to K^- f_0 \nu_{\tau}$}

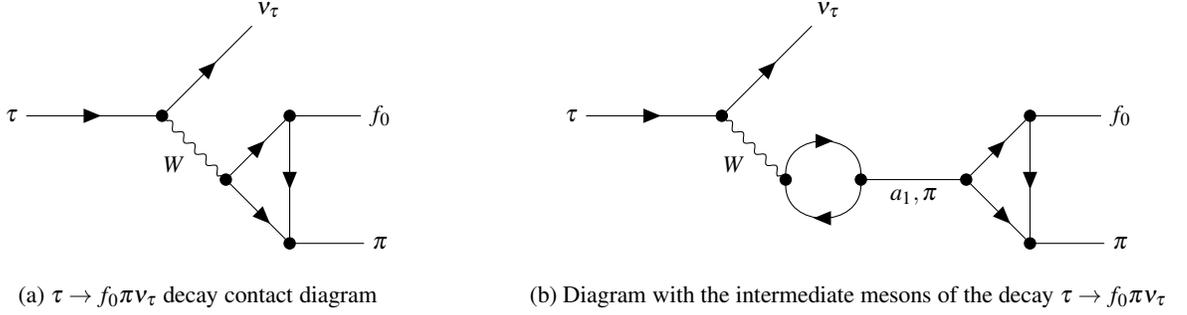
\begin{figure*}[t]
 \centering
  \begin{subfigure}{0.5\textwidth}
   \centering
    \begin{tikzpicture}
     \begin{feynman}
      \vertex (a) {\(\tau\)};
      \vertex [dot, right=2cm of a] (b){};
      \vertex [above right=2cm of b] (c) {\(\nu_{\tau}\)};
      \vertex [dot, below right=1.2cm of b] (d) {};
      \vertex [dot, above right=1.2cm of d] (e) {};
      \vertex [dot, below right=1.2cm of d] (h) {};
      \vertex [right=1.2cm of e] (f) {\(f_0\)};
      \vertex [right=1.2cm of h] (k) {\(\pi\)}; 
      \diagram* {
         (a) -- [fermion] (b),
         (b) -- [fermion] (c),
         (b) -- [boson, edge label'=\(W\)] (d),
         (d) -- [fermion] (e),  
         (e) -- [fermion] (h),
         (h) -- [anti fermion] (d),
         (e) -- [] (f),
         (h) -- [] (k),
      };
     \end{feynman}
    \end{tikzpicture}
   \caption{$\tau \to f_0 \pi \nu_\tau$ decay contact diagram}
  \end{subfigure}%
 \centering
 \begin{subfigure}{0.5\textwidth}
  \centering
   \begin{tikzpicture}
    \begin{feynman}
      \vertex (a) {\(\tau\)};
      \vertex [dot, right=2cm of a] (b){};
      \vertex [above right=2cm of b] (c) {\(\nu_{\tau}\)};
      \vertex [dot, below right=1.2cm of b] (d) {};
      \vertex [dot, right=1cm of d] (l) {};
      \vertex [dot, right=1.4cm of l] (g) {};
      \vertex [dot, above right=1.2cm of g] (e) {};
      \vertex [dot, below right=1.2cm of g] (h) {};
      \vertex [right=1.2cm of e] (f) {\(f_0\)};
      \vertex [right=1.2cm of h] (k) {\(\pi\)}; 
      \diagram* {
         (a) -- [fermion] (b),
         (b) -- [fermion] (c),
         (b) -- [boson, edge label'=\(W\)] (d),
         (d) -- [fermion, inner sep=1pt, half left] (l),
         (l) -- [fermion, inner sep=1pt, half left] (d),
         (l) -- [edge label'=\({a_1, \pi} \)] (g),
         (g) -- [fermion] (e),  
         (e) -- [fermion] (h),
         (h) -- [anti fermion] (g),
         (e) -- [] (f),
         (h) -- [] (k),
      };
     \end{feynman}
    \end{tikzpicture}
   \caption{Diagram with the intermediate mesons of the decay $\tau \to f_0 \pi \nu_\tau$}
  \end{subfigure}%
 \caption{Diagrams contributing to the decay $\tau \to f_0 \pi \nu_\tau$.}
 \label{diagram2}
\end{figure*}

    The diagrams of the process $\tau \to \pi^- f_0 \nu_{\tau}$ are presented in Fig.~\ref{diagram2}.

    The amplitude of the decay $\tau \to \pi^- f_0(500) \nu_{\tau}$ in the NJL model takes the following form:
    \begin{eqnarray}
    \label{amplitude1}
        \mathcal{M} = -i G_F V_{ud} \sqrt{Z_\pi} \cos\bar{\phi} L_\mu \left\{\left[g^{\mu\nu} + \left(g^{\mu\nu}\left(q^2 - 6m_u^2\right) - q^\mu q^\nu\right)BW_{a_1}\right] \left(p_\pi - p_f\right)_\nu - 4 m_u^2 q^\mu BW_{\pi}\right\},
    \end{eqnarray}
    where $G_F$ is the Fermi constant, $V_{ud}$ is an element of the Cabibbo-Kobayashi-Maskawa matrix, $L_\mu$ is the weak current, $q$ is the momentum of the intermediate state, and $p_\pi$ and $p_f$ are the momenta of the final mesons.

    The amplitude of this process consists of the axial vector channel including the contact contribution and the contribution from the intermediate $a_1$ meson and pseudoscalar channel containing the contribution from the pion. The intermediate states are described by the Breit-Wigner propagator:
    \begin{eqnarray}
        BW_{meson} = \frac{1}{M_{meson}^2 - q^2 - i\sqrt{q^2}\Gamma_{meson}},
    \end{eqnarray}
    where $M$ and $\Gamma$ are the meson masses and widths. For the calculations we use $M_{a_1}=1230$ MeV, $M_{f_0(500)}=550$ MeV, $\Gamma_{a_1}=400$ MeV and $\Gamma_{f_0(500)}=500$ MeV~\cite{ParticleDataGroup:2024cfk}.  
    
    The decay $\tau \to \pi^- f_0(980) \nu_{\tau}$ can be obtained from the decay $\tau \to \pi^- f_0(500) \nu_{\tau}$ by the replacement of $\cos\bar{\phi}$ by $\sin\bar{\phi}$ in the amplitude (\ref{amplitude1}). 

    The amplitude of the decay $\tau \to K^- f_0(500) \nu_{\tau}$ in the NJL model differs from the previous ones in that it contains the strange mesons:
    \begin{eqnarray}
    \label{amplitude4}
        \mathcal{M} & = & -\frac{i}{2} G_f V_{us} Z_K L_\mu \left\{\left[\frac{g_{{f_0}_u}}{g_K}\cos\bar{\phi} + \sqrt{2} \frac{{g_{{f_0}_s}}}{g_K}\sin\bar{\phi}\right]\left[g^{\mu\nu} + \left(g^{\mu\nu}\left(q^2 - \frac{3}{2}(m_s + m_u)^2\right) - q^\mu q^\nu\right)\sin\alpha BW_{K_1(1270)} \right.\right.\nonumber\\
        && \left. + \left(g^{\mu\nu}\left(q^2 - \frac{3}{2}(m_s + m_u)^2\right) - q^\mu q^\nu\right)\cos\alpha BW_{K_1(1400)}\right] \left(p_K - p_f\right)_\nu \nonumber\\
        && \left. - 2\left[(2m_u - m_s)\frac{{g_{{f_0}_u}}}{g_K}\cos\bar{\phi} + \sqrt{2} (2m_s - m_u) \frac{g_{{f_0}_s}}{g_K}\sin\bar{\phi}\right] (m_u + m_s) q^\mu BW_K \right\}.
    \end{eqnarray}

    The decay $\tau \to K^- f_0(980) \nu_{\tau}$ can be obtained from the decay $\tau \to K^- f_0(500) \nu_{\tau}$ by the replacement of $\cos\bar{\phi} \to \sin\bar{\phi}$ and $\sin\bar{\phi} \to -\cos\bar{\phi}$ in the amplitude (\ref{amplitude4}).

    The theoretical estimations of the branching fractions of these decays are shown in Table~\ref{tab_width}.
    The uncertainty of numerical calculations within the model is estimated at $\pm15\%$ \cite{Volkov:2005kw,Volkov:2022ukj,Volkov:2022nxp}.

    \begin{table}[h!]
    \begin{center}
    \begin{tabular}{cc}
    \hline
    Decays &  $Br({\tau \to SP \nu_\tau})$  \\
    \hline
    $\tau \to f_0(500)\pi \nu_\tau$  & $(5.06 \pm 0.76) \times 10^{-2}$  \\
    $\tau \to f_0(980)\pi \nu_\tau$  & $(2.53 \pm 0.37) \times 10^{-4}$  \\
    $\tau \to f_0(500) K \nu_\tau$   & $(9.70 \pm 1.45) \times 10^{-5}$  \\
    $\tau \to f_0(980) K \nu_\tau$	 & $(5.42 \pm 0.81) \times 10^{-6}$   \\
    \hline
    \end{tabular}
    \end{center}
    \caption{Branching fractions of the decays of the $\tau$ lepton with production of scalar mesons}
    \label{tab_width}
\end{table}

\section{Decay $\tau \to \pi^-2\pi^0 \nu_{\tau}$ taking into account the state $f_0(500)$}

\begin{figure*}[t]
 \centering
  \begin{subfigure}{0.5\textwidth}
   \centering
    \begin{tikzpicture}
     \begin{feynman}
      \vertex (a) {\(\tau\)};
      \vertex [dot, right=2cm of a] (b){};
      \vertex [above right=2cm of b] (c) {\(\nu_{\tau}\)};
      \vertex [dot, below right=1.2cm of b] (d) {};
      \vertex [dot, above right=1.2cm of d] (e) {};
      \vertex [dot, below right=1.2cm of d] (h) {};
      \vertex [dot, right=1.0cm of e] (f) {};
      \vertex [dot, above right=1.0cm of f] (n) {};  
      \vertex [dot, below right=1.0cm of f] (m) {};   
      \vertex [right=1.2cm of n] (l) {\(\ \pi \)}; 
      \vertex [right=1.2cm of m] (s) {\(\pi\)};  
      \vertex [right=1.4cm of h] (k) {\(\pi\)}; 
      \diagram* {
         (a) -- [fermion] (b),
         (b) -- [fermion] (c),
         (b) -- [boson, edge label'=\(W\)] (d),
         (d) -- [fermion] (e),  
         (e) -- [fermion] (h),
         (d) -- [anti fermion] (h),
         (e) -- [edge label'=\({ \rho, f_0 } \)] (f),
         (f) -- [fermion] (n),
         (n) -- [fermion] (m),
         (f) -- [anti fermion] (m), 
         (h) -- [] (k),
         (n) -- [] (l),
	 (m) -- [] (s),
      };
     \end{feynman}
    \end{tikzpicture}
  \end{subfigure}%
 \caption{Quark diagram for the $\tau \to 3\pi \nu_\tau$ decay contact channel.}
 \label{diagram3}
\end{figure*}
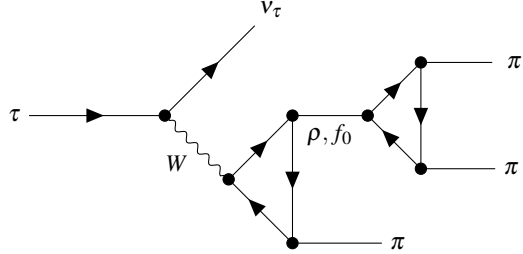%

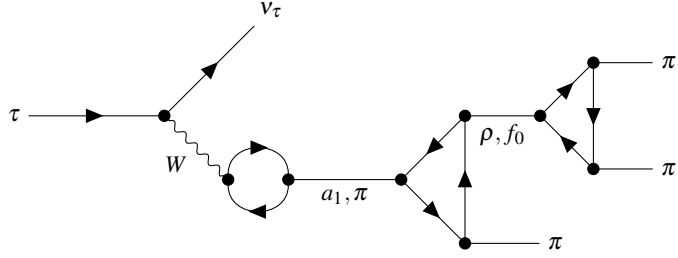
\begin{figure*}[t]
 \centering
 \centering
 \begin{subfigure}{0.5\textwidth}
  \centering
   \begin{tikzpicture}
    \begin{feynman}
      \vertex (a) {\(\tau\)};
      \vertex [dot, right=2cm of a] (b){};
      \vertex [above right=2cm of b] (c) {\(\nu_{\tau}\)};
      \vertex [dot, below right=1.2cm of b] (d) {};
      \vertex [dot, right=0.8cm of d] (l) {};
      \vertex [dot, right=1.5cm of l] (g) {};
      \vertex [dot, above right=1.2cm of g] (e) {};
      \vertex [dot, below right=1.2cm of g] (h) {};      
      \vertex [dot, right=1.0cm of e] (f) {};
      \vertex [dot, above right=1.0cm of f] (n) {};
      \vertex [dot, below right=1.0cm of f] (m) {};
      \vertex [right=1.0cm of n] (s) {\( \pi \)};
      \vertex [right=1.0cm of m] (r) {\( \pi \)};
      \vertex [right=1.2cm of h] (k) {\( \pi \)}; 
      \diagram* {
         (a) -- [fermion] (b),
         (b) -- [fermion] (c),
         (b) -- [boson, edge label'=\(W\)] (d),
         (d) -- [fermion, inner sep=1pt, half left] (l),
         (l) -- [fermion, inner sep=1pt, half left] (d),
         (l) -- [edge label'=\({ a_1, \pi } \)] (g),
         (g) -- [fermion] (h),  
         (h) -- [fermion] (e),
         (e) -- [fermion] (g),      
         (e) -- [edge label'=\( {\rho, f_0}\)] (f),
         (f) -- [fermion] (n),
         (n) -- [fermion] (m),
         (m) -- [fermion] (f),
         (h) -- [] (k), 
         (n) -- [] (s),
         (m) -- [] (r),
      };
     \end{feynman}
    \end{tikzpicture}
  \end{subfigure}%
 \caption{Diagram with intermediate mesons describing the decay $\tau \to 3 \pi \nu_\tau$.}
 \label{diagram4}
\end{figure*}%

The process $\tau \to 3\pi \nu_{\tau}$ is the most probable decay of the $\tau$ lepton into three mesons. The amplitude of the decay $\tau \to \pi^-2\pi^0 \nu_{\tau}$ consists of the contributions of the intermediate axial vector and pseudoscalar channels with the vertices $a_1\rho\pi$, $a_1 f_0\pi$, $\pi\rho\pi$ and $\pi f_0\pi$. The diagrams describing the decay $\tau \to \pi^-2\pi^0 \nu_{\tau}$ are presented in Figs. \ref{diagram3} and \ref{diagram4}. For all channels, $\rho(770)$ and $f_0(500)$ are taken into account as the second intermediate states. 
The total amplitude of this decay can be written in the following form:
\begin{eqnarray}
\label{amplitude2}
\mathcal{M}(\tau \to \pi^- \pi^0 \pi^0 \nu_\tau) & = & -i F_{\pi} G_{F} V_{ud} L_{\mu} \left\{ \mathcal{M}_{a_1\rho} + \mathcal{M}_{\pi\rho} + \mathcal{M}_{a_1f_0} + \mathcal{M}_{\pi f_0} \right\}^{\mu},
\end{eqnarray}
where the hadron part of the amplitude is presented in curly brackets. $F_\pi=m_u/g_\pi$ is the weak pion decay constant. Here the first and third terms describe the axial vector channel, i. e. the channel with the axial vector meson in the first intermediate state. The second and the fourth terms describe the pseudoscalar channel. The first two terms describe the contributions containing the $\rho$ meson in the second intermediate state. The last two terms describe the contributions with the scalar meson $f_0(500)$ in the second intermediate state. The expressions for the hadron parts of the amplitude take the form:
\begin{eqnarray}
\mathcal{M}_{a_1\rho}^{\mu} & = & Z_\pi g^2_\rho
 \left[ g_{\mu\nu} + \left((q^{2} - 6m_u^2)g^{\mu\nu} - q^{\mu}q^{\nu} \right) BW_{a_1} \right] BW_{\rho} (p_{\pi^-} - p^{(1)}_{\pi^0})_\nu + \left(p^{(1)}_{\pi^0} \leftrightarrow p^{(2)}_{\pi^0} \right), 
\end{eqnarray}
\begin{eqnarray}
\mathcal{M}_{\pi\rho}^{\mu} & = & g^2_\rho BW_{\pi} q_\mu {\left(q+ p^{(2)}_{\pi^-} \right)}_\nu {\left(p_{\pi^-} - p^{(1)}_{\pi^0} \right)}_\nu BW_{\rho} + \left(p^{(1)}_{\pi^0} \leftrightarrow p^{(2)}_{\pi^0} \right), 
\end{eqnarray}
\begin{eqnarray}
\mathcal{M}_{a_1f_0}^{\mu} & = & 4 Z_\pi^2 g^2_{{f_0}_u} \cos^2\bar{\phi}
 \left[ g_{\mu\nu} + \left((q^{2} - 6m_u^2)g^{\mu\nu} - q^{\mu}q^{\nu} \right) BW_{a_1} \right] BW_{f_0} (p_{\pi^-} - p^{(1)}_{\pi^0} - p^{(2)}_{\pi^0})_\nu + \left(p^{(1)}_{\pi^0} \leftrightarrow p^{(2)}_{\pi^0} \right),
\end{eqnarray}
\begin{eqnarray}
 \mathcal{M}_{\pi f_0}^{\mu} & = & 16 Z_\pi^2 m_u g^2_{{f_0}_u} \cos^2\bar{\phi} q^\mu BW_{\pi} BW_{f_0} + \left(p^{(1)}_{\pi^0} \leftrightarrow p^{(2)}_{\pi^0} \right).
\end{eqnarray}
where $p_{\pi^-}$, $p^{(1)}_{\pi^0}$ and $p^{(2)}_{\pi^0}$ are the momenta of the final charged and neutral pions. The contact contributions are included into the axial vector channels. They are described by the first terms in the square brackets.

The interference between the scalar and $\rho$ meson contributions play an important role.

As a result, the branching fraction of this decay is in agreement with the experimental data within the uncertainties:
\begin{eqnarray}
    Br(\tau \to \pi^- \pi^0 \pi^0 \nu_\tau)_{NJL} & = & (9.15 \pm 0.45) \%, \nonumber\\
    Br(\tau \to \pi^- \pi^0 \pi^0 \nu_\tau)_{exp} & = & (9.26 \pm 0.1) \% \textrm{ \cite{ParticleDataGroup:2024cfk}}.
\end{eqnarray}

The values of the separate contributions are given in Table~\ref{tab_width2}.

\begin{table}[h!]
\begin{center}
\begin{tabular}{ccc}
\hline
Channels of the decay &  $B$ fractions (\%) & CLEO II \cite{CLEO:1999rzk} \\
\hline
$\rho\pi$ 	   & 60.0                &  $68.11$\\
$f_0(500)\pi$  & 12.8                &   $16.18 \pm 3.85 $ \\
\hline
\end{tabular}
\end{center}
\caption{The separate contributions of the channels with the vector and the scalar mesons. The values of the branching fractions are normalized to the full branching fraction of the decay $\tau \to \pi^-\pi^0\pi^0 \nu_\tau$. The sum of the channels is not equal to $100 \%$ because of the interference between the contributions.}
\label{tab_width2}
\end{table} 

\section{Conclusion}
In this paper, the $\tau$ lepton decays with the production of mesons $f_0(\pi,K)$ and neutrinos are described within the NJL quark model. The contact channels and channels with the intermediate axial-vector and pseudoscalar channels are considered. It is shown that the contact and axial-vector channels play a dominant role. The performed calculations shows a fairly large value of the decay width $\tau \to f_0(500) \pi \nu_\tau$. Unfortunately, experimental measurement of the partial width of this decay is difficult due to the wide decay width $f_0(500) \to \pi \pi$  \cite{Volkov:1998ax}. Therefore, the role of the decay $\tau \to f_0(500) \pi \nu_\tau$ in the process $\tau \to 3\pi \nu_\tau$ should be investigated. 
Indeed, experimental data demonstrate the significant presence of an intermediate scalar meson in the process $\tau \to \pi^-\pi^0\pi^0 \nu_\tau$ \cite{CLEO:1999rzk}. The performed calculations for the decay $\tau \to \pi^-\pi^0\pi^0 \nu_\tau$ in the NJL model are in satisfactory agreement with experiment (see Table \ref{tab_width2}). In this process, the interference between the channels with the intermediate $\rho$ and $f_0(500)$ mesons plays an important role. The agreement of theoretical results with experiment for the process $\tau \to \pi^-\pi^0\pi^0 \nu_\tau$ allows us to hope for the reliability of the obtained predictions for the decays $\tau \to f_0(500)(\pi,K) \nu_\tau$ and $\tau \to f_0(980)(\pi,K) \nu_\tau$. Note that when calculating the decay $\tau \to \pi^-\pi^0\pi^0 \nu_\tau$, the contributions from the "box" diagram were not taken into account, since in the works \cite{K:2023kgj,Volkov:2023pmy} it was shown that these contributions are negligibly small.

The obtained results in the present work for the decay $\tau \to f_0(980) \pi \nu_\tau$ are in satisfactory agreement with the results of calculations carried out in \cite{Dai:2018rra} within the model using the mechanism with a triangular singularity.

Similar calculations can also be performed for the decays $\tau \to a_0 \eta \nu_\tau$ and $\tau \to a_0 K \nu_\tau$. As a result, for the partial widths of the decays we obtain $Br(\tau \to a_0 \eta \nu_\tau) = 4.3 \times 10^{-5}$, $Br(\tau \to a_0 K \nu_\tau) = 5 \times 10^{-6}$. The corresponding amplitudes of these decays are given in Appendix \ref{appendix}. However, since the tetraquark component, which is not taken into account in the NJL model, probably plays an important role in describing the quark structure of the meson $a_0$, here we can only count on the qualitative nature of the obtained results for the decays with $a_0$ meson production.
    
\appendix
\section{Amplitudes of the decays $\tau \to a_0 \eta \nu_\tau$ and $\tau \to a_0 K \nu_\tau$}
\label{appendix}

The calculations carried out in the NJL model lead to the following amplitude of the decay $\tau \to a_0 \eta \nu_\tau$
    \begin{eqnarray}
        \mathcal{M} = -i G_F V_{ud} F_\pi \sqrt{Z_\pi} \sin\bar{\theta} L_\mu \left\{\left[g^{\mu\nu} + \left(g^{\mu\nu}\left(q^2 - 6m_u^2\right) - q^\mu q^\nu\right)BW_{a_1}\right] \left(p_\eta - p_{a_0}\right)_\nu - 4 m^2_u q^\mu BW_{\pi}\right\},
    \end{eqnarray}
where $q=p_{a_0}+p_\eta$. 

The amplitude of the process $\tau \to a_0 K^- \nu_\tau$ in the NJL model takes the form
    \begin{eqnarray}
        \mathcal{M} & = & -\frac{i}{2} G_F V_{us} Z_K \frac{g_{{f_0}_u}}{g_K} L_\mu \left\{\left[g^{\mu\nu} + \left(g^{\mu\nu}\left(q^2 - \frac{3}{2}(m_u + m_s)^2\right) - q^\mu q^\nu\right)\left(\sin\alpha BW_{K_1(1270)} + \cos\alpha BW_{K_1(1400)}\right) \right]\right. \nonumber\\
        && \left. \times \left(p_K - p_{a_0}\right)_\nu - 2 (m_u + m_s) (2m_u - m_s) q^\mu BW_{K}\right\},
    \end{eqnarray}
where $q=p_{a_0}+p_K$. 

These amplitudes include the contact contributions and the contributions containing the intermediate axial vector and pseudoscalar states.

\subsection*{Acknowledgements}
    The authors thank prof. A. B. Arbuzov for useful discussions.

\end{document}